\documentclass[12pt, a4paper]{article}

\usepackage[utf8]{inputenc}
\usepackage[T1]{fontenc}
\usepackage[english]{babel}
\usepackage{mathptmx} 
\usepackage{amsmath}
\usepackage{amsfonts}
\usepackage{bm}      
\usepackage[left=2.5cm, right=2.5cm, top=2.5cm, bottom=2.5cm]{geometry}
\usepackage{setspace}
\usepackage{booktabs}
\usepackage{array}
\usepackage{tabularx}
\usepackage{adjustbox}
\usepackage{longtable}
\usepackage{float} 
\usepackage{ragged2e} 
\newcolumntype{L}[1]{>{\RaggedRight\arraybackslash}p{#1}}
\usepackage{graphicx}
\usepackage{xcolor}
\usepackage[longnamesfirst]{natbib}
\usepackage{hyperref}
\hypersetup{
  colorlinks=true,
  linkcolor=blue,
  citecolor=blue,
  urlcolor=cyan
}

\begin{document}
\onehalfspacing

\title{\textbf{The Crisis Simulator for Bolivia (KISr-p): An Empirically Grounded Modeling Framework}}
\author{Ricardo Alonzo Fernández Salguero}
\date{\today}

\maketitle

\begin{abstract}
\justify
This document presents a detailed technical report of the ``Crisis Simulator for Bolivia (KISr-p),'' a quarterly stochastic model designed to evaluate the impact of various macroeconomic policy strategies in an environment of high uncertainty and structural constraints. Unlike standard general equilibrium frameworks, this simulator is grounded in the consolidated empirical findings from a vast collection of meta-analyses, adopting a theoretical architecture of a Keynesian Intertemporal Synthesis (KIS) with a Constant Elasticity of Substitution (KIS-CES) production function. The calibration of each model block—real, fiscal, monetary, external, labor, and distributional—is described, justifying key parameters with quantitative evidence on the hierarchy of fiscal multipliers \citep{GechertRannenberg2018}, the complementarity of production factors \citep{GechertHavranekIrsovaKolcunova2022}, monopsony power in the labor market \citep{SokolovaSorensen2021}, and the dynamics of exchange rate and interest rate pass-through. It is demonstrated how the model integrates these regularities to generate non-linear dynamics, such as the dependence of multipliers on the state of the business cycle and the asymmetric impact of shocks. The results of a set of simulated policy scenarios are presented, illustrating the trade-offs between fiscal adjustment, external financing, debt restructuring, and structural reforms, such as the aggressive restructuring of spending. The analysis concludes that a pragmatic policy approach, which prioritizes the composition of spending over its aggregate level and recognizes institutional frictions, offers superior macroeconomic and welfare outcomes compared to dogmatic solutions.
\end{abstract}

\section{Introduction}

Contemporary macroeconomics is at a crossroads, marked by a proliferation of empirical results that are often contradictory and context-dependent. This ``cacophony of results,'' as described by \citet{Doucouliagos2013}, has hindered the formulation of economic policies based on a robust consensus, allowing debates to persist on an ideological plane. Economic liberalism, popularized through frameworks like the Washington Consensus, has traditionally advocated for deregulation, privatization, and market flexibility as paths to prosperity \citep{DeHaan2006, FernandezSalguero2025CW}. However, aggregated evidence, once purged of publication bias and methodological heterogeneity, often challenges these dogmas, revealing a world with significant frictions, market power, and a crucial role for the state and institutions \citep{FernandezSalguero2025KIS_OpenEconomy}.

Faced with this challenge, quantitative synthesis through meta-analysis has established itself as an indispensable tool for distilling robust signals from empirical noise \citep{Stanley2001, Irsova2024}. By systematically analyzing thousands of estimates from primary studies, meta-analyses have revealed empirical regularities that demand a recalibration of our theoretical frameworks. Findings such as the superiority of investment spending multipliers in recessions \citep{GechertRannenberg2018}, the null effect of the minimum wage on employment due to monopsony power \citep{DoucouliagosStanley2009, SokolovaSorensen2021, FernandezSalguero2025Labor}, and the elasticity of substitution between capital and labor consistently below unity \citep{GechertHavranekIrsovaKolcunova2022}, are not anomalies but structural features of modern economies.

This document presents the ``Crisis Simulator for Bolivia (KISr-p),'' a macroeconomic policy laboratory that takes this aggregated evidence as its foundation. Instead of starting from pure theoretical axioms, the simulator adopts the Keynesian Intertemporal Synthesis with CES production (KIS-CES) framework, a structure that reconciles intertemporal optimization with Keynesian frictions, and calibrates it directly with the consolidated parameters from the meta-analysis literature \citep{FernandezSalguero2025KIS_EN}. The objective is to build a bridge between the vast empirical evidence and the practical modeling of policies, offering a tool to analyze complex scenarios in a small, open economy subject to external shocks and structural constraints, as is the case of Bolivia \citep{FernandezSalguero2025BoliviaIP, FernandezSalguero2025MESCP}.

The analysis was executed with the \textit{Bolivia Crisis Simulator KISr-p} \citep{fernandez_salguero_2025_bolivia}, available on \href{https://github.com/RAFS20/Bolivia-Crisis-Simulator-KISr-p/tree/main}{\texttt{GitHub/RAFS20}}.

\section{Simulator Architecture and Methodology}

The ``Crisis Simulator for Bolivia (KISr-p)'' is a macroeconomic policy laboratory that operates as a stochastic quarterly time-series model. Its purpose is not point forecasting, but the comparative risk analysis of different strategies over a medium-term horizon (40 quarters). The core methodology is simulation, where for each scenario, hundreds of possible trajectories (\texttt{N\_PATHS}) are generated, each subject to a different sequence of random shocks, allowing for the construction of distributions of future outcomes.

The model is not a computable general equilibrium (CGE) system but a recursive structural model with path-dependence. The state variables in a quarter \texttt{t} (debt, reserves, exchange rate gap) determine the conditions for quarter \texttt{t+1}, influencing multipliers and risk premia. This structure captures crucial non-linear dynamics, such as tipping points where a fall in reserves can trigger a currency crisis, or where an increase in debt can lead to a spiral of higher risk premia. The architecture is organized into interconnected blocks:

\begin{enumerate}
    \item \textbf{Real and Fiscal Block:} Determines GDP through an aggregate demand equation that incorporates the effects of fiscal policy (current spending, investment, transfers), global shocks, and the impact of monetary policy. The calibration of fiscal multipliers is state-dependent (boom, recession, crisis), reflecting the evidence from \citet{GechertRannenberg2018}. The fiscal side is closed with an endogenously counter-cyclical primary balance rule and a revenue function sensitive to the business cycle and inflation.
    
    \item \textbf{Prices, Monetary, and External Block:} Models inflation with components of inertia, the output gap, and a pass-through channel from the parallel exchange rate, the magnitude of which is non-linear and dependent on reserves \citep{Iorngurum2025a}. Monetary policy follows a Taylor rule, but its effect on the real economy is modest, consistent with the corrected evidence of \citet{FernandezSalguero2025Monetary}. The external sector is modeled through a simplified balance of payments, where the capital account responds to the sovereign risk premium.
    
    \item \textbf{Debt and Risk Block:} The debt/GDP dynamic evolves according to the deficit, financial cost, and growth. The sovereign risk premium is a key variable, modeled endogenously as a non-linear function of debt, reserves, the exchange rate gap, and political instability, drawing lessons from the literature on debt crises and austerity \citep{FernandezSalguero2025Austerity}.
    
    \item \textbf{Labor, Distributional, and Welfare Block:} Includes a labor market with monopsony power, consistent with \citet{SokolovaSorensen2021} and \citet{FernandezSalguero2025Informality}. This allows social policies (\texttt{TR}) to affect not only demand but also inequality (Gini) and a proxy for human capital/health. A synthetic welfare index is calculated for ranking scenarios.
\end{enumerate}

This modular design allows for the simulation of a wide range of policies, from traditional spending cuts to reforms such as the implementation of a land value tax (LVT), whose feasibility is analyzed in \citet{FernandezSalguero2025IVS}, or the interaction with multilateral organization programs, whose impacts are a matter of debate \citep{FernandezSalguero2025IMF}.

\section{Foundations of the Real and Fiscal Block}

The core of the simulator lies in its fiscal block, designed to capture the most robust empirical regularities on the effectiveness of spending and taxes. Neoclassical theory traditionally posits that tax cuts are more effective than spending and that public spending crowds out private investment. The aggregated evidence, however, tells a different story.

The meta-analysis by \citet{Gechert2015} is fundamental to the model's calibration. It finds that public spending multipliers are systematically larger than those for tax cuts (a difference of 0.3-0.4 points) and establishes a clear hierarchy: the multiplier for public investment (infrastructure, R\&D) is approximately 0.5 points higher than that for current spending. This is because investment increases long-term productive capacity, an effect corroborated by \citet{BomLigthart2014}. The analysis for Bolivia by \citet{FernandezSalguero2025BoliviaIP} confirms this hierarchy, showing a strong crowding-in effect of infrastructure investment. The simulator implements this evidence by distinguishing between Current Spending (GC), Public Investment (GI), and Transfers (TR), with multipliers that respect the relationship $m_{GI} > m_{TR} > m_{GC}$.

The most influential finding post-2008 crisis is the business cycle dependency of fiscal effectiveness. The meta-analysis by \citet{GechertRannenberg2018} demonstrates that multipliers are significantly more potent in recessions (increasing by 0.7 to 0.9 points). The simulator captures this non-linearity through a regime-switching mechanism. In each quarter, the economy is classified as ``Boom,'' ``Recession,'' or ``Crisis'' based on the output gap and reserves. Depending on the regime, multipliers are drawn from a different distribution, being highest in a ``Crisis.'' This implies that the same austerity policy can have a devastating effect in a recession, a central result in the critique of such policies \citep{FernandezSalguero2025Austerity}. Table \ref{tab:multiplicadores} summarizes the base calibration.

\begin{table}[H]
\centering
\caption{Calibration of Mean Quarterly Fiscal Multipliers by Regime}
\label{tab:multiplicadores}
\begin{tabularx}{\textwidth}{L{4.5cm} >{\centering\arraybackslash}X >{\centering\arraybackslash}X >{\centering\arraybackslash}X}
\toprule
\textbf{Type of Spending} & \textbf{Boom} & \textbf{Recession} & \textbf{Crisis} \\
\midrule
Current Spending (GC) & 0.045 & 0.125 & 0.213 \\
Transfers (TR) & 0.075 & 0.200 & 0.263 \\
Public Investment (GI) & 0.175 & 0.388 & 0.525 \\
\bottomrule
\end{tabularx}
\justify
\footnotesize{\textit{Note:} The values represent the mean quarterly impact of a fiscal impulse of 1\% of GDP. The calibration follows the hierarchy and cycle dependency from \citet{Gechert2015} and \citet{GechertRannenberg2018}. The values are drawn from a t-Student distribution in each simulation.}
\end{table}

\section{Monetary, Exchange Rate, and External Sector Block}

In a small, open economy like Bolivia, the monetary and external channels are crucial. The simulator models them with a pragmatic approach, anchored in the evidence that the power of conventional monetary policy is limited. The meta-analysis by \citet{FernandezSalguero2025Monetary}, which reviews a vast literature, finds a publication bias that exaggerates the effects of monetary policy. Once corrected, the peak impact of a 100 bps interest rate shock on output is reduced from -1.0\% to a modest -0.25\%. Consequently, the simulator incorporates a transmission channel from the interest rate to aggregate demand that is deliberately weak.

The exchange rate channel is, in practice, more relevant. The model distinguishes between an official exchange rate (\texttt{S\_off}) and a parallel one (\texttt{S\_par}), generating an endogenous gap. The dynamics of the gap respond to balance of payments pressures (reflected in the level of reserves), the risk premium, and social unrest. Inflation, in turn, is sensitive to movements in the parallel exchange rate through a non-linear pass-through mechanism: the pass-through is greater when reserves are low, capturing the erosion of credibility. The evidence from \citet{Iorngurum2025a} and \citet{VelickovskiPugh2011} on incomplete and asymmetric pass-through justifies this modeling.

The external sector is modeled with a simplified balance of payments. The current account (CA) responds to the real exchange rate and the business cycle, and the capital account (KA) to the risk premium. The change in reserves (\texttt{Res}) is the net result. A crucial element is the critical reserve threshold (\texttt{RES\_CRIT}). If reserves fall below this level, a ``Crisis'' regime is triggered, raising multipliers and risk, and increasing the probability of an exchange rate realignment. This mechanism is consistent with the literature on financial crises \citep{FernandezSalguero2025Austerity}. Table \ref{tab:canales_externos} summarizes the main channels.

\begin{table}[H]
\centering
\caption{Main Transmission Channels of the Monetary and External Block}
\label{tab:canales_externos}
\begin{tabularx}{\textwidth}{L{4cm} X}
\toprule
\textbf{Channel} & \textbf{Mechanism and Empirical Justification} \\
\midrule
\textbf{Interest Rate} & An increase in the policy rate has a small contractionary effect on GDP and inflation. Transmission is attenuated by dollarization and the exchange rate gap. Calibration based on \citet{FernandezSalguero2025Monetary}. \\
\addlinespace
\textbf{Exchange Rate Pass-through} & Inflation responds to changes in the \textit{parallel} exchange rate. The pass-through coefficient ($\phi_t$) is non-linear: it increases as reserves fall below a critical threshold. Consistent with \citet{Iorngurum2025a}. \\
\addlinespace
\textbf{Sovereign Risk Premium} & The spread (RP) is endogenous. It increases with debt/GDP and the gap; it decreases with reserves and institutional quality. This generates a feedback loop, as documented in \citet{FernandezSalguero2025Austerity}. \\
\addlinespace
\textbf{Capital Account} & Capital flows are volatile and pro-cyclical. They respond negatively to the risk premium (capital flight) and social unrest. Policies like temporary capital flow management (CFM) can moderate this volatility. \\
\bottomrule
\end{tabularx}
\end{table}

\section{Labor Market, Inequality, and Welfare}

The KIS-CES simulator incorporates frictions and market power in the labor block, following the evidence from meta-analyses. The finding of a practically null effect of the minimum wage on employment \citep{DoucouliagosStanley2009, FernandezSalguero2025Labor} is consistent with a market with monopsony power, a widespread characteristic according to \citet{SokolovaSorensen2021}. The model implements a ``monopsony-lite'' version, where employment (\texttt{Emp}) co-evolves from demand and supply functions, allowing fiscal shocks to have a direct impact on employment, in line with the theoretical framework of \citet{FernandezSalguero2025Informality}.

The model also tracks distributional and social variables. Inequality (Gini) evolves in response to unemployment and fiscal progressivity. Social transfers (\texttt{TR}) reduce the Gini coefficient. In turn, inequality has feedback effects: the meta-analyses of \citet{CaprettiTonni2025} and \citet{Cipollina2018} find a negative relationship between inequality and growth. In the simulator, a higher Gini increases social unrest (\texttt{Unrest}) and reduces potential growth through its impact on human capital/health (\texttt{Health}).

The health/human capital index (\texttt{Health}) improves with employment and social investment, and deteriorates with inequality. This captures the findings of meta-analyses such as those by \citet{Kondo2009} and \citet{Shimonovich2024}, which document a robust negative association between inequality and health. The inclusion of these variables allows for a more holistic evaluation of policies. The final welfare index aggregates these dimensions to provide a comparative ranking, as summarized in Table \ref{tab:bienestar}.

\begin{table}[H]
\centering
\caption{Components of the Synthetic Welfare Index}
\label{tab:bienestar}
\begin{tabularx}{\textwidth}{L{4cm} >{\centering\arraybackslash}p{1.5cm} X}
\toprule
\textbf{Component} & \textbf{Weight} & \textbf{Justification and Link to Literature} \\
\midrule
GDP Level & +0.30 & Proxy for aggregate income and consumption. \\
Employment Level & +0.15 & Captures the labor dimension of welfare. \\
Health Index & +0.10 & Represents human capital. Its link to inequality is based on \citet{Kondo2009}. \\
Inflation & -0.15 & High inflation erodes purchasing power. \\
Debt/GDP & -0.10 & High debt implies vulnerability. \\
Exchange Rate Gap & -0.08 & Signal of imbalances and distortions. \\
Social Unrest & -0.07 & Proxy for political instability. \\
Gini Coefficient & -0.05 & Inequality has direct social costs. \\
\bottomrule
\end{tabularx}
\end{table}

\section{Mathematical Formalization of the Stochastic Model}
\label{sec:modelo_matematico}

The model simulates the trajectory of a small, open economy with external constraints and a dual exchange rate market. It is formulated in discrete time, with a period $t$ representing a quarter. The system is defined by a state vector that evolves stochastically over a time horizon of $T=40$ quarters. The structural blocks that govern the model's dynamics are detailed below. Table \ref{tab:variables} presents a summary of the main notation.

\subsection{Real Sector: Output Dynamics and its Determinants}

Aggregate output, $Y_t$, evolves as a function of fiscal policy, external shocks, and the endogenous dynamics of the system. The output gap, $g_t$, is defined as the percentage deviation of observed output from its potential level, $Y_t^P$:
\begin{equation}
g_t = \frac{Y_t - Y_t^P}{Y_t^P}
\end{equation}
The economy's behavior is classified into one of three regimes, $\mathcal{R}_t \in \{\text{Boom, Recession, Crisis}\}$, determined by the output gap and the level of international reserves, $R_t$, relative to a critical threshold $R_{crit}$:
\begin{equation}
\mathcal{R}_t = 
\begin{cases} 
\text{Crisis}   & \text{if } (R_t \le R_{crit}) \lor (\mathcal{B}_t > \mathcal{B}_{crisis}) \\
\text{Recession} & \text{if } (g_t < 0) \land (\mathcal{R}_t \neq \text{Crisis}) \\
\text{Boom}     & \text{if } (g_t \ge 0) \land (\mathcal{R}_t \neq \text{Crisis})
\end{cases}
\end{equation}
where $\mathcal{B}_t = S_t^{par} / S_t^{off}$ is the exchange rate gap.

The real output growth rate, $\Delta \ln Y_{t+1}$, responds to fiscal impulses, shocks, and the monetary policy channel:
\begin{equation}
\Delta \ln Y_{t+1} = -\mu_t^{GC} \Delta G_t^C + \mu_t^{GI} \Delta G_t^I + \mu_t^{TR} \Delta T_t^R + z_t + m_t + \epsilon_t^d
\label{eq:crecimiento_y}
\end{equation}
where $\Delta G_t^C$, $\Delta G_t^I$, and $\Delta T_t^R$ represent changes in current spending, public investment, and transfers as a proportion of GDP. The fiscal multipliers, $\mu_t^{(\cdot)}$, are stochastic and depend on the economic regime $\mathcal{R}_t$, following a theoretical hierarchy where $\mu_t^{GI} > \mu_t^{TR} > \mu_t^{GC}$ \cite{Gechert2015, AlarconGambarte2020}. The magnitude of the multipliers increases significantly during recessions and crises, in line with empirical evidence highlighting the greater effectiveness of fiscal stimulus when there are idle resources \cite{BlanchardLeigh2013, Christiano2011, GechertRannenberg2018}. The term $z_t$ is an autoregressive global shock, $m_t$ captures the modest impact of the interest rate channel, and $\epsilon_t^d$ is an idiosyncratic demand shock with fat tails, distributed according to a t-Student.

Potential output, $Y_t^P$, evolves at a base rate, $g^P$, but is adjusted by public capital accumulation, $K_t^P$, and by distributional factors. Public capital accumulation follows:
\begin{equation}
K_{t+1}^P = (1 - \delta_p) K_t^P + G_t^I Y_t
\end{equation}
where $\delta_p$ is the quarterly depreciation rate. The dynamics of potential output are expressed as:
\begin{equation}
\Delta \ln Y_{t+1}^P = g^P + \alpha_p \frac{\Delta K_{t+1}^P}{Y_t} + \beta_{gini} (Gini_t - \overline{Gini})
\end{equation}
where $\alpha_p$ is the marginal productivity of public capital \cite{BomLigthart2014} and the last term introduces a channel through which higher income inequality ($Gini_t$) can erode long-term growth \cite{CaprettiTonni2025, Cipollina2018}.

\subsection{Prices and Inflation}
Quarterly inflation, $\pi_t$, follows a process that combines inertia, a Phillips curve, an exchange rate pass-through channel, and an anchor to the inflation target, $\pi^*$:
\begin{equation}
\pi_t = \rho_{\pi} \pi_{t-1} + \kappa g_t + \phi(R_t) \Delta \ln S_t^{par} + (1 - \rho_{\pi})(\pi^*) + \epsilon_t^{\pi}
\label{eq:inflacion}
\end{equation}
The pass-through from the parallel exchange rate, $\Delta \ln S_t^{par}$, to prices is non-linear, mediated by the function $\phi(R_t)$, which increases as liquid reserves, $R_t$, fall below the critical threshold $R_{crit}$. This non-linearity captures the de-anchoring of inflation expectations in contexts of foreign currency scarcity \cite{Iorngurum2025a, VelickovskiPugh2011}. The pass-through function has a logistic form:
\begin{equation}
\phi(R_t) = \frac{\bar{\phi}}{1 + \exp\left(k_{\phi} \left( \frac{R_t}{R_{crit}} - 1 \right)\right)}
\end{equation}
where $\bar{\phi}$ is the maximum pass-through and $k_{\phi}$ modulates its sensitivity to reserve scarcity.

\subsection{Public Finances and Fiscal Rule}
Public revenues as a fraction of GDP, $\tau_t$, are endogenous and depend on economic activity, inflation, and institutional credibility, $Cred_t$:
\begin{equation}
\tau_t = \bar{\tau} + \beta_g g_t + \beta_{\pi} \pi_t + \beta_{cred} (Cred_t - \overline{Cred})
\end{equation}
The primary balance target, $pd_t^*$, follows a convergence path from an initial deficit to a medium-term surplus, $pd_{target}$. Additionally, it incorporates a counter-cyclical fiscal rule that seeks to stabilize debt dynamics \cite{Heimberger2023b}:
\begin{equation}
pd_t^* = (1 - a_t) pd_0 + a_t pd_{target} - \gamma (r_t^{eff} - \hat{g}_t) b_t
\end{equation}
where $a_t$ is a logistic convergence factor, $b_t$ is the debt-to-GDP ratio, $r_t^{eff}$ is the effective interest rate on debt, and $\hat{g}_t$ is the nominal GDP growth rate. The coefficient $\gamma$ is dependent on the regime $\mathcal{R}_t$, being lower in crises to avoid excessive pro-cyclicality \cite{BlanchardLeigh2013}.

The observed primary deficit, $pd_t$, results from adjusting the target for discretionary spending and investment policies:
\begin{equation}
pd_t = pd_t^* + (\Delta G_t^C + \Delta G_t^I + \Delta T_t^R) - (\tau_t - \bar{\tau})
\end{equation}

\subsection{Debt, Interest Rates, and Sovereign Risk}
The law of motion for the debt-to-GDP ratio, $b_t$, is:
\begin{equation}
b_{t+1} = \frac{1+r_t^{eff}}{1+\hat{g}_t}b_t + pd_t + \lambda^{FX}_t b_t \left( \frac{S_{t+1}^{off}}{S_t^{off}} - 1 \right)
\label{eq:ley_mov_deuda}
\end{equation}
The last term captures the valuation effect of foreign currency-denominated debt, whose proportion is $\lambda^{FX}_t$, in the event of a devaluation of the official exchange rate, $S_t^{off}$ \cite{Krugman1999}.

The quarterly country risk, $rp_t$, is a function of the debt level, reserves, the exchange rate gap, social unrest ($Unrest_t$), and institutional quality:
\begin{equation}
rp_t = \rho_{rp} rp_{t-1} + (1 - \rho_{rp}) \left[ f(b_t) + \beta_R R_t + \beta_{\mathcal{B}} (\mathcal{B}_t-1) + \beta_U Unrest_t + \beta_{IFI} \mathbb{I}_{IFI} + \boldsymbol{\beta_I'Z_I} \right]
\end{equation}
where $f(b_t)$ is a logistic function increasing in the debt level \cite{MendozaYue2012}, $\mathbb{I}_{IFI}$ is an indicator variable for the presence of a program with International Financial Institutions (IFIs), which can reduce perceived risk \cite{Vreeland2003}, and $\boldsymbol{Z_I}$ is a vector of institutional quality variables (strength of fiscal rules, transparency). The model can incorporate debt restructuring events (PSI/OSI) that reduce the stock of debt and/or the interest burden, generating a discrete drop in $rp_t$, although these events may have reputational costs \cite{FernandezSalguero2025IMF}. The effective interest rate, $r_t^{eff}$, gradually adjusts towards the market rate ($r_t^{market} = r_f + rp_t$), reflecting the average maturity of the debt portfolio.

\subsection{External Sector and Exchange Markets}
The change in reserves as a proportion of GDP, $\Delta R_t$, is governed by the balance of payments:
\begin{equation}
\Delta R_t = CA_t + KA_t
\end{equation}
The current account, $CA_t$, depends on the real exchange rate, the output gap, and fiscal spending components, while the capital account, $KA_t$, is sensitive to country risk, social unrest, and the exchange rate gap:
\begin{align}
CA_t &= \bar{CA} + \eta_{CA,S} \Delta \ln S_t^{par} + \eta_{CA,g} g_t + \boldsymbol{\eta_{CA,G}' \Delta G_t} + \epsilon_t^{CA} \\
KA_t &= \eta_{KA,rp} rp_t + \eta_{KA,U} Unrest_t - f_{KA,\mathcal{B}}(\mathcal{B}_t) + \epsilon_t^{KA}
\end{align}
where $f_{KA,\mathcal{B}}(\cdot)$ is a non-linear function that captures the acceleration of capital flight as the exchange rate gap widens.

The exchange rate system is dual. The official exchange rate, $S_t^{off}$, can be fixed, follow a crawling peg, or undergo a discrete realignment (devaluation) if reserves fall below $R_{crit}$, a mechanism similar to first-generation crisis models \cite{FloodGarber1984, KaminskyReinhart1999}. The parallel exchange rate, $S_t^{par}$, floats, and its dynamics determine the gap $\mathcal{B}_t$:
\begin{equation}
\Delta \ln \mathcal{B}_{t+1} = \alpha_{\mathcal{B},R} \max\left(0, \frac{R_{crit}}{R_{t+1}}-1\right) + \alpha_{\mathcal{B},rp} rp_t + \alpha_{\mathcal{B},U} Unrest_t - \alpha_{\mathcal{B},cred} Cred_t + \epsilon_t^{\mathcal{B}}
\end{equation}
This equation reflects that pressure on the parallel market increases with reserve scarcity, country risk, and social unrest, while greater policy credibility mitigates it.

\subsection{Labor Market, Distribution, and Welfare}
The level of employment, $E_t$ (as a fraction of the active population), is determined by a combination of demand and supply factors, approximating a market with monopsony power \cite{SokolovaSorensen2021, FernandezSalguero2025Informality}.
\begin{equation}
E_{t+1} = (1-\omega_E) E_t + \omega_E (\phi_d g_t - \phi_r (r_t^{eff}-\hat{g}_t)) + (1-\omega_E) (\eta_s \Delta w_t) + \epsilon_t^E
\end{equation}
where $w_t$ is the real wage. Income inequality, measured by the Gini coefficient, responds negatively to progressive social transfers and positively to unemployment:
\begin{equation}
Gini_{t+1} = Gini_t + \beta_{G,TR} \Delta T_t^R + \beta_{G,E} (1 - E_{t+1}) + \epsilon_t^{Gini}
\end{equation}
Finally, a social welfare index, $\mathcal{W}_t$, is constructed as a weighted aggregate of macroeconomic and social variables, normalized for comparison.
\begin{equation}
\mathcal{W}_t = \sum_{i} w_i \cdot f_i(X_{i,t})
\end{equation}
where $X_{i,t}$ is a vector of key indicators (level of $Y_t$, $E_t$, health, $\pi_t$, $b_t$, $\mathcal{B}_t$, $Unrest_t$, $Gini_t$) and $w_i$ are the weights assigned to each component. This index is an ad hoc tool for the comparative ranking of policy scenarios, not a cardinal measure of social utility. Its inclusion reflects the growing concern for the distributional and social effects of adjustment policies, often overlooked in traditional economic paradigms \cite{Stiglitz2005, FernandezSalguero2025Austerity}.

\begin{longtable}{@{} p{0.2\textwidth} p{\dimexpr 0.8\textwidth - 2\tabcolsep} @{}}
\caption{Notation of the Main Variables and Parameters of the Model} \label{tab:variables} \\
\toprule
\textbf{Symbol} & \textbf{Description} \\
\midrule
\endfirsthead
\multicolumn{2}{c}%
{{\bfseries \tablename\ \thetable{} -- continued from previous page}} \\
\toprule
\textbf{Symbol} & \textbf{Description} \\
\midrule
\endhead
\midrule
\multicolumn{2}{r}{{Continued on next page}} \\ 
\endfoot
\bottomrule
\endlastfoot
$Y_t$ & Real Gross Domestic Product (GDP) in quarter $t$. \\
$Y_t^P$ & Potential GDP. \\
$g_t$ & Output gap. \\
$\mathcal{R}_t$ & Economic regime (Boom, Recession, Crisis). \\
$\pi_t$ & Quarterly inflation rate. \\
$\pi^*$ & Long-term inflation target. \\
$b_t$ & Public debt to GDP ratio. \\
$pd_t$ & Primary deficit as a fraction of GDP. \\
$r_t^{eff}$ & Quarterly effective interest rate on debt. \\
$\hat{g}_t$ & Nominal GDP growth rate. \\
$rp_t$ & Quarterly sovereign risk premium. \\
$R_t$ & Net international reserves as a fraction of GDP. \\
$R_{crit}$ & Critical level of reserves. \\
$S_t^{off}$ & Official nominal exchange rate. \\
$S_t^{par}$ & Parallel nominal exchange rate. \\
$\mathcal{B}_t$ & Exchange rate gap ($S_t^{par} / S_t^{off}$). \\
$CA_t, KA_t$ & Current account and capital account balances. \\
$E_t$ & Employment level (fraction of the active population). \\
$w_t$ & Real wage (index). \\
$Gini_t$ & Gini coefficient for income inequality. \\
$Unrest_t$ & Social unrest index. \\
$Cred_t$ & Policy credibility index. \\
$\mathcal{W}_t$ & Synthetic social welfare index. \\
$\mu_t^{(\cdot)}$ & Fiscal multipliers (GC: Current Spending, GI: Investment, TR: Transfers). \\
$\lambda_t^{FX}$ & Proportion of debt denominated in foreign currency. \\
$\epsilon_t^{(\cdot)}$ & Stochastic shocks with zero mean. \\
\end{longtable}

\section{Summary of Macroeconomic Simulation Results}

A set of simulations (300 paths per strategy) was executed over a 40-quarter horizon (10 years) to evaluate the performance of 29 different macroeconomic strategies. The main objective is to compare the outcomes in terms of debt sustainability, economic growth, and social welfare. The key indicators at the end of the period (Quarter 40) are the median Debt/GDP ratio (\texttt{Debt\_med}), the median GDP level (\texttt{GDP\_med}) relative to the starting point, and a composite social welfare metric (\texttt{Welfare\_med}).

The results show a very significant divergence among the different families of strategies. On one hand, shock fiscal adjustment policies, especially those accompanied by an initial devaluation without a debt restructuring framework, lead to the worst outcomes, with a debt ratio exceeding 200\% of GDP and a drastic fall in welfare. On the other hand, the most successful strategies are those that combine an aggressive fiscal restructuring with debt reprofiling, financing from international financial institutions (IFIs), and, crucially, the implementation of structural reforms like a land value tax (LVT). These strategies not only manage to contain but also reduce the debt ratio to below 110\% of GDP, while maintaining a higher level of economic activity and achieving welfare levels that are double those of conventional strategies.

Between these two extremes lies a group of conventional strategies (gradual adjustments, standard IMF-type programs, and partial restructurings) that, while avoiding collapse, fail to stabilize debt dynamics, ending the period with ratios in the 160\% to 170\% of GDP range. Within this group, those that include debt relief (PSI/OSI) and IFI financing show a markedly superior performance to those based solely on fiscal adjustment, managing to reduce the final debt by approximately 30 percentage points of GDP and substantially improving welfare.

\subsection{Comparative Table of Strategies (Median Results at T=40)}

The following table presents a summary of the median results for all evaluated strategies at the end of the simulation horizon (Quarter 40). The strategies are ordered by their Debt/GDP level, from lowest to highest.

\begin{longtable}{l c c c}
\caption{Median Results at Quarter 40 by Strategy.} \label{tab:resumen_completo} \\
\toprule
\textbf{Strategy} & \textbf{Median GDP} & \textbf{Median Debt (\%)} & \textbf{Median Welfare} \\
\midrule
\endfirsthead
\toprule
\textbf{Strategy} & \textbf{Median GDP} & \textbf{Median Debt (\%)} & \textbf{Median Welfare} \\
\midrule
\endhead
\bottomrule
\endfoot
AggRecomp\_GI+TR\_modDebt\_MKT\_LVT & 98.35 & 106.25 & 20.63 \\
AggRecomp\_GI+TR\_modDebt\_IFI\_LVT & 99.24 & 107.46 & 21.44 \\
AggRecomp\_GI+TR\_lowDebt\_IFI\_LVT & 97.62 & 108.08 & 20.90 \\
AggRecomp\_GI+TR\_noDebt\_LVT & 97.93 & 108.24 & 20.27 \\
Aggressive-Recomp+PSI+OSI+IFI & 98.61 & 133.25 & 17.35 \\
Balanced-Recomp+PSI+OSI+IFI & 97.69 & 135.62 & 17.84 \\
PSI30+IFI1.5 & 98.70 & 150.15 & 16.35 \\
Minimal-Recomp & 100.21 & 157.62 & 14.78 \\
Gradual-1.5 & 98.58 & 159.26 & 12.77 \\
Gradual-3 & 98.76 & 159.58 & 12.74 \\
LMO-LCY-Shift & 96.43 & 161.39 & 13.20 \\
Balanced-Recomp & 99.22 & 161.50 & 15.52 \\
IMF-2025-base+SoftMonetary & 99.18 & 162.01 & 14.66 \\
SCD+IFI2 & 97.01 & 162.07 & 14.69 \\
IMF-no-IFI & 97.35 & 162.32 & 13.53 \\
IMF-2025-base & 98.23 & 163.26 & 14.77 \\
Swap+4pp & 98.59 & 164.49 & 13.50 \\
Loans-Only & 97.48 & 165.22 & 13.80 \\
Aggressive-Recomp & 97.25 & 165.55 & 14.95 \\
AggRecomp\_GI+TR\_lowDebt\_IFI\_noLVT & 98.13 & 167.12 & 13.95 \\
Shock-8 & 95.64 & 171.03 & 11.93 \\
Shock-12 & 93.48 & 171.26 & 11.19 \\
Crawl15\%+CFM8q & 98.69 & 172.95 & 14.22 \\
AggRecomp\_GI+TR\_hiDebt\_IFI\_Deval & 98.18 & 196.42 & 11.44 \\
AggRecomp\_GI+TR\_hiDebt\_MKT\_Deval & 97.69 & 197.65 & 11.18 \\
AggRecomp\_GI+TR\_noDebt\_noDeval\_noLVT & 97.27 & 199.19 & 10.68 \\
AggRecomp\_GI+TR\_noDebt\_Deval\_noLVT & 96.87 & 199.79 & 10.24 \\
Shock-6+Deval & 96.66 & 201.38 & 9.53 \\
Shock-6+Deval+DebtMKT & 94.90 & 204.18 & 8.74 \\
\multicolumn{4}{p{\textwidth}}{\footnotesize{\textit{Note:} Strategy names are coded. "AggRecomp" denotes an aggressive spending recomposition towards Investment (GI) and Transfers (TR). "Balanced-Recomp" is a more moderate version. "LVT" indicates a Land Value Tax reform. "PSI/OSI" refers to debt restructuring with Private/Official Sector Involvement. "IFI" means financing from International Financial Institutions.}}
\end{longtable}

\subsection{Ranking of Strategies by Highest Welfare}
This table highlights the five strategies with the highest median welfare level at the end of the period, while also maintaining a relatively high GDP level. It is observed that strategies combining aggressive fiscal spending and structural restructuring (LVT) clearly dominate, followed by comprehensive packages that include debt restructuring (PSI/OSI) and IFI support.

\begin{table}[ht!]
\centering
\caption{Top 5 Strategies by Welfare and GDP at T=40.}
\label{tab:top_welfare}
\begin{tabular}{l c c c c}
\toprule
\textbf{Strategy} & \textbf{Median GDP} & \textbf{Debt (\%)} & \textbf{Welfare} & \textbf{Reserves} \\
\midrule
AggRecomp\_GI+TR\_modDebt\_IFI\_LVT & 99.24 & 107.46 & 21.44 & 3.97 \\
AggRecomp\_GI+TR\_lowDebt\_IFI\_LVT & 97.62 & 108.08 & 20.90 & 3.17 \\
AggRecomp\_GI+TR\_modDebt\_MKT\_LVT & 98.35 & 106.25 & 20.63 & 3.99 \\
AggRecomp\_GI+TR\_noDebt\_LVT & 97.93 & 108.24 & 20.27 & 2.00 \\
Balanced-Recomp+PSI+OSI+IFI & 97.69 & 135.62 & 17.84 & 21.21 \\
\bottomrule
\end{tabular}
\end{table}

\subsection{Evolution of Key Indicators for Selected Strategies}
To illustrate the divergence in trajectories, the following table shows the evolution of debt and welfare for four representative strategies: a high-performing one (with LVT), a well-performing one (with PSI/OSI), a conventional one (IMF), and a low-performing one (shock with devaluation).

\begin{table}[ht!]
\centering
\caption{Median Trajectories of Debt and Welfare for Representative Strategies.}
\label{tab:trayectorias}
\begin{tabular}{l c c c c c c}
\toprule
& \multicolumn{3}{c}{\textbf{Median Debt (\%)}} & \multicolumn{3}{c}{\textbf{Median Welfare}} \\
\cmidrule(lr){2-4} \cmidrule(lr){5-7}
\textbf{Strategy} & \textbf{T=0} & \textbf{T=16} & \textbf{T=40} & \textbf{T=0} & \textbf{T=16} & \textbf{T=40} \\
\midrule
AggRecomp\_GI+TR\_modDebt\_MKT\_LVT & 110.0 & 91.49 & 106.25 & 14.46 & 25.36 & 20.63 \\
Balanced-Recomp+PSI+OSI+IFI & 110.0 & 83.59 & 135.62 & 14.44 & 25.72 & 17.84 \\
IMF-base+SoftMonetary & 110.0 & 121.49 & 162.01 & 14.13 & 21.30 & 14.66 \\
Shock-6+Deval & 110.0 & 163.20 & 201.38 & 14.33 & 15.98 & 9.53 \\
\bottomrule
\end{tabular}
\end{table}

\subsection{Graphical Analysis of Results}

To complement the tabular analysis, the following figures visualize the key relationships between the main outcome variables at the end of the simulation period (T=40). These graphs help identify performance clusters and visualize the inherent trade-offs of the different policies.

\begin{figure}[ht!]
\centering
\includegraphics[width=0.9\textwidth]{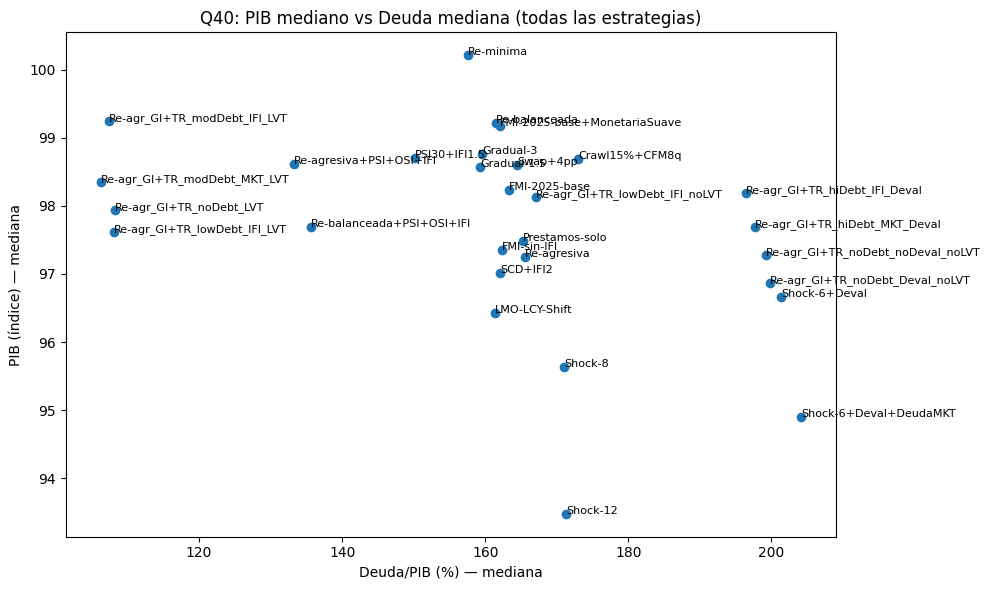}
\caption{Relationship between GDP and Debt/GDP at the end of the period (T=40).}
\label{fig:pib_vs_deuda}
\end{figure}

\textbf{Figure 1} shows the relationship between the median GDP level and the median Debt/GDP ratio. The upper-left quadrant represents the most desirable outcome: high GDP with low debt. A clear segmentation of strategies is observed. An elite group, composed almost exclusively of strategies implementing a Land Value Tax (\texttt{...LVT}), is located in the top-left corner, achieving the lowest debt ratios (between 106\% and 110\%) without sacrificing the level of economic activity. At the opposite extreme, in the bottom-right corner, are the shock and fiscal adjustment strategies without debt relief (\texttt{Shock-6+Deval}, \texttt{AggRecomp\_...\_noLVT}), which result in explosive debt (over 195\%) and a comparatively lower GDP. The vast majority of conventional strategies (\texttt{IMF-base}, \texttt{Gradual}, etc.) are clustered in an intermediate zone, with high debt levels (155\%-175\%) and a GDP that, while relatively high, does not compensate for fiscal unsustainability.

\begin{figure}[ht!]
\centering
\includegraphics[width=0.9\textwidth]{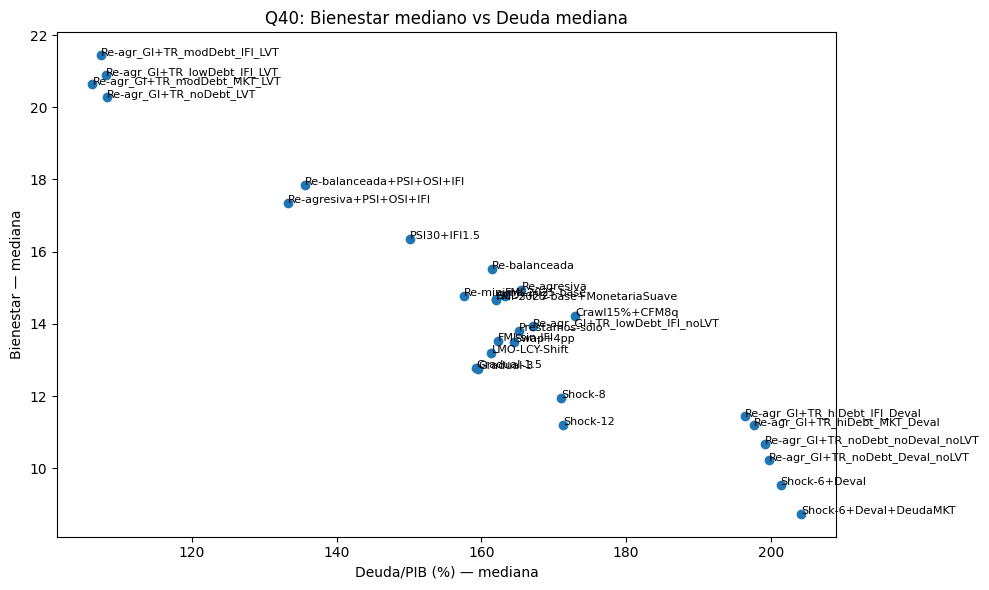}
\caption{Relationship between Welfare and Debt/GDP at the end of the period (T=40).}
\label{fig:welfare_vs_deuda}
\end{figure}

\textbf{Figure 2} graphs the relationship between median welfare and debt. The negative correlation is even more pronounced than in the case of GDP. The results are stratified into four clear levels:
\begin{enumerate}
    \item \textbf{Exceptional Performance (top left):} Strategies with LVT achieve a much higher level of welfare (above 20), demonstrating that fiscal solvency achieved through structural reforms has a direct and massive impact on social welfare.
    \item \textbf{Good Performance (upper center):} Strategies that include significant debt relief (\texttt{...PSI+OSI+IFI}) are situated in a second tier, with high welfare (17-18) and debt contained around 135\%. This underscores the importance of debt restructuring for social recovery.
    \item \textbf{Mediocre Performance (lower center):} The cluster of conventional strategies shows considerably lower welfare levels (12-15), indicating that the persistence of a high debt burden limits the ability to generate welfare, even with a stable GDP.
    \item \textbf{Poor Performance (bottom right):} Pure shock and adjustment strategies without debt relief cause welfare to collapse to levels below 12, confirming that severe austerity with rising debt is the worst combination for society.
\end{enumerate}

\begin{figure}[ht!]
\centering
\includegraphics[width=\textwidth]{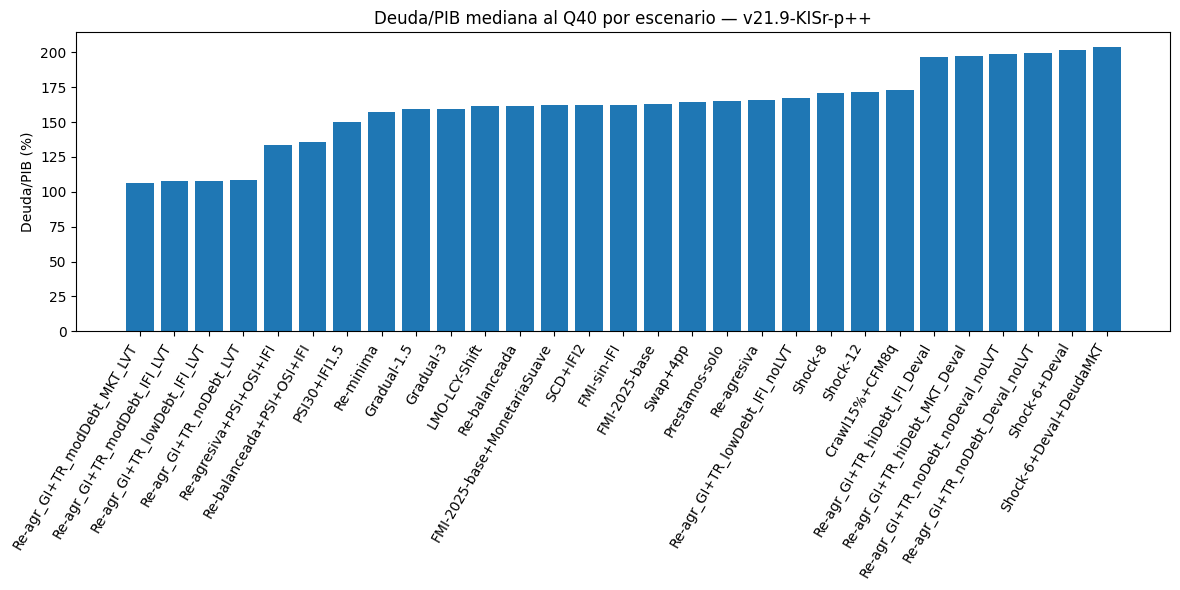}
\caption{Median Debt/GDP Ratio at Q40 ordered by scenario.}
\label{fig:deuda_bar_chart}
\end{figure}

\textbf{Figure 3} offers an unequivocal ranking of all strategies based on their median debt outcome at the end of the horizon. This bar chart visually confirms the stratification observed in the scatter plots. Clear steps grouping families of policies are visible. The first group, with debt contained below 110\%, corresponds to policies with LVT. It is followed by a second step with debt relief strategies (PSI/OSI) that stabilize around 135\%-150\%. Next, a long plateau of conventional policies is observed, which fail to break below the 157\% debt floor. Finally, the graph shows a rapid ascent to unsustainable levels for strategies of severe adjustment without restructuring. Overall, this figure compellingly summarizes the main finding: without deep structural reforms or substantial debt relief, conventional policies are not sufficient to resolve the debt crisis.

\section{Conclusions}

The Crisis Simulator for Bolivia (KISr-p) represents an effort to build a policy analysis framework that is theoretically coherent, computationally robust, and, above all, disciplined by the vast empirical evidence consolidated through meta-analyses. By adopting a KIS-CES architecture that incorporates heterogeneity, frictions, and market power, the model is able to replicate the key empirical regularities that define modern macroeconomics.

The conclusions emerging from the analysis of the simulated scenarios challenge conventional narratives. First, the simulator's evidence is clear: \textbf{the composition of fiscal adjustment matters more than its aggregate magnitude}. Strategies based on indiscriminate spending cuts tend to be contractionary and worsen debt sustainability. Conversely, strategies that reallocate resources from current spending to public investment in high-multiplier physical and human capital can be expansionary. This empirically validates the central pillar of evidence on the importance of productive investment as a dual engine of demand and supply.

Second, the model underscores that \textbf{simple solutions are insufficient for complex problems}. Neither fiscal adjustment alone, nor external financing, nor debt restructuring are panaceas. The most robust results are achieved through comprehensive policy packages that combine an \textit{intelligent} fiscal adjustment (aggressive spending restructuring) with debt burden relief and structural reforms that strengthen institutional anchors. Pragmatism, based on a nuanced reading of the evidence, overcomes ideological dogmatism.

The modeling exercise demonstrates the urgency of integrating the findings of quantitative synthesis into the practice of economic policy. Ignoring the cycle-dependency of multipliers, the existence of monopsony power in the labor market, or the complementary nature of public and private capital is not a harmless simplification; it is an error that leads to suboptimal and harmful policy recommendations. The KIS-CES simulator offers a platform for a more informed policy dialogue, where decisions are based not on a single study or a single theory, but on the aggregated weight of decades of empirical research.

\end{document}